# A Constraint Network Based Approach to Memory Layout Optimization*

G. Chen and M. Kandemir
Computer Science and Engineering Department
The Pennsylvania State University, University Park
PA 16802, USA
{guilchen, kandemir}@cse.psu.edu

M. Karakoy
Department of Computing
Imperial College
London, SW7 2AZ, UK
m.karakoy@ic.ac.uk

**Abstract**

While loop restructuring based code optimization for array intensive applications has been successful in the past, it has several problems such as the requirement of checking dependences (legality issues) and transformation of all of the array references within the loop body indiscriminately (while some of the references can benefit from the transformation, others may not). As a result, data transformations, i.e., transformations that modify memory layout of array data instead of loop structure have been proposed. One of the problems associated with data transformations is the difficulty of selecting a memory layout for an array that is acceptable to the entire program (not just to a single loop). In this paper, we formulate the problem of determining the memory layouts of arrays as a constraint network, and explore several methods of solution in a systematic way. Our experiments provide strong support in favor of employing constraint processing, and point out future research directions.

## 1. Introduction

Data locality of array-based computations has been an exciting research area for the last decade or so. Most of the prior proposals to the problem are based on *loop transformations* [8][11][10][16], i.e., modifying the order of loop iterations to make data access pattern more cache friendly. Loop transformations have several key advantages that make them appealing to compiler writers and users alike. First, there is a comprehensive theory behind them developed over the years [16] and supported through several commercial implementations. Second, they are proven to be effective in enhancing both temporal and spatial locality. Third, and maybe most importantly, transformation of a loop nest is independent of transformations of other nests in the same application code. In other words, its impact is localized to the nest in question. Consequently, for each loop nest, one can use the best loop transformation from the data locality perspective without worrying about the interactions between neighboring loop nests.

---

*This work is supported in part by NSF Career Award #0093082.

However, recent research has revealed several drawbacks of loop transformations such as the requirement of checking dependences (legality issues) and transformation all of the references within the loop body indiscriminately (while some of the references can benefit from the transformation, others may not). As a result, *data transformations* [1][6][12], i.e., transformations that modify memory layout of array data instead of loop structure have been proposed. While data transformations do not have the problems associated with loop transformations, it has proven to be difficult to implement robust data transformation frameworks, mainly because of the fact that a memory layout modification affects all references to the array in question in all the loop nests of the application (i.e., localized optimization is not possible). In other words, its impact is global and difficult to capture. Therefore, prior efforts mainly concentrated on heuristic approaches whose results could not have been validated in formal terms.

In this work, we focus on data transformations from a different perspective, and treat them within the paradigm of *constraint processing*. In more specific terms, we formulate the problem of determining the memory layouts of arrays for a given application as a *constraint network* [3], and explore several methods of solution in a systematic way. In doing so, our ultimate goal is two-fold. First, we want to show that constraint processing provides an attractive approach to implement a data transformation framework. Second, using the solutions returned by this framework, we want to take a fresh look at previously proposed heuristic solutions to the problem, and check how they compare to our constraint network based approach. This paper reports on our experience with this constraint processing based solution, and presents an empirical evaluation. Specifically, we designed and implemented a constraint network specialized for solving

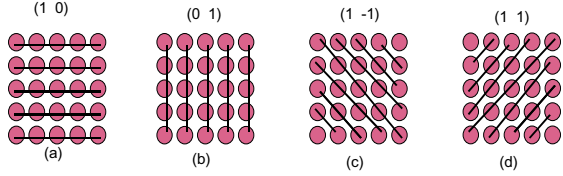

**Figure 1 Different memory layouts for a two-dimensional array (data space) and the corresponding hyperplane vectors.**

```
for (i1=0;i1<N;i1++)
  for(i2=0;i2<N;i2++)
    …Q1[i1+i2][i2]…Q2[i1+i2][i1]…
```

**Figure 2. An example loop nest.**

memory layout problems. Our experiments provide strong support in favor of employing constraint processing, and point out future research directions.

The rest of this paper is organized as follows. The next section discusses our memory layout representation based on linear algebra. Section 3 describes our constraint network and gives a formal definition of the problem. Section 4 discusses backtracking and backjumping based solutions to the problem. An experimental evaluation of our approach is presented in Section 5, and we conclude the paper in Section 6.

## 2. Hyperplane-Based Memory Layout Representation

Our memory layout representation is based on linear algebra and makes use of spatial locality in memory space. In a k-dimensional space, a hyperplane is defined as a set of tuples $(x_1\ x_2\ \dots\ x_k)$ that satisfy the equation $x_1 y_1 + x_2 y_2 + \dots + x_k y_k = c$, where $(y_1\ y_2\ \dots\ y_k)$ represents hyperplane coefficients (also called hyperplane vector) and c is the hyperplane constant. Note that $(y_1\ y_2\ \dots y_k)$ represents a hyperplane family, each member of which has a different constant (c value) [7]. Two points represented by column vectors, $d_1$ and $d_2$, are said to belong to the same hyperplane if:

$(y_1\ y_2 \dots y_k) \bullet d_1 = (y_1\ y_2 \dots y_k) \bullet d_2$,

where $\bullet$ denotes point multiplication.[1] As an example, in a two-dimensional data space, the hyperplane vector (1 0) indicates that two array elements belong to the same hyperplane as long as they have the same value for the row index.

Let us now focus explicitly on a two-dimensional space (an extension to higher dimensional spaces will be discussed later). Note that, a hyperplane family can be used to partially describe the memory layout of an array. For example, if we do not care about the relative order of hyperplanes, we can use hyperplane vector (1 0) to denote

---
[1] The point multiplication of two vectors $(x_1\ x_2\ x_3\ \dots\ x_k)$ and $(y_1\ y_2\ y_3\ \dots\ y_k)^T$ is $x_1 y_1 + x_2 y_2 + x_3 y_3 + \dots + x_k y_k$.

row-major memory layout in a two-dimensional space. This is because two array elements $d_1 = (d_{11}\ d_{12})^T$ and $d_2 = (d_{21}\ d_{22})^T$ belong to the same row if and only if:

$(1\ 0) \bullet (d_{11}\ d_{12})^T = (1\ 0) \bullet (d_{21}\ d_{22})^T$;

that is, if and only if $d_{11} = d_{21}$. In other words, as long as the two array elements have the same row index, they belong to the same hyperplane, which corresponds to a row in a two-dimensional array. Note that, while all the rows of the array have the same hyperplane vector, their hyperplane coefficients (c values) are different from each other (in fact, a hyperplane coefficient in this example corresponds to the row number). Figure 1(a) depicts such a row-major layout and shows hyperplanes explicitly. Figures 1(b) through 1(d), on the other hand, illustrate different memory layouts and give their hyperplane vectors. Let us briefly concentrate on the diagonal layout shown in Figure 1(c). In this layout, the two data elements $d_1 = (d_{11}\ d_{12})^T$ and $d_2 = (d_{21}\ d_{22})^T$ are stored in the same diagonal if and only if $(1\ -1) \bullet (d_{11}\ d_{12})^T = (1\ -1) \bullet (d_{21}\ d_{22})^T$, which means $d_{11} - d_{12} = d_{21} - d_{22}$. For example, $(5\ 3)^T$ and $(7\ 5)^T$ are stored in the same diagonal, whereas $(5\ 3)^T$ and $(5\ 4)^T$ are on two different diagonals. Note that, there are other possible diagonal layouts as well. For example, hyperplane vectors (1 -2) and (2 -1) also indicate diagonal layouts (which are different from (1 -1)).

An important point to note here is that, in order to have good data locality, data access pattern should be along the same direction with the hyperplane vector. Let us focus on a row-major memory layout for illustrative purposes (see Figure 1(a)). In order to have good spatial locality, two successive loop iterations, denoted by I and $I^n$ (not that in a nest with multiple loops I and $I^n$ are vectors), should access the array elements $d_1$ and $d_2$ such that $(1\ 0) \bullet d_1 = (1\ 0) \bullet d_2$.

In this paper, however, we are interested in determining the best memory layout for a given data access pattern. Therefore, our problem is to choose a hyperplane vector $(y_1\ y_2)$ such that:

$(y_1\ y_2) \bullet d_1 = (y_1\ y_2) \bullet d_2$,

assuming that $d_1$ and $d_2$ are the array elements accessed by I and $I^n$. As an example, consider the nested loop shown in Figure 2. In this nest, we have two references to two different arrays ($Q_1$ and $Q_2$). Assuming that $I = (i_1\ i_2)^T$ and $I^n = (i_1\ i_2+1)^T$ are two successive loop iterations that do not cross loop bounds, for array $Q_1$, we should find a hyperplane vector $(y_1 \ y_2)$ representing its memory layout such that the following equality should be satisfied:

$$(y_1 \ y_2) \bullet (i_1+i_2 \ i_2)^T = (y_1 \ y_2) \bullet (i_1+i_2+1 \ i_2+1)^T,$$

which means $(y_1 \ y_2) = (1 \ -1)$, i.e., the diagonal layout.[2] Similarly, for array $Q_2$, we need to satisfy:

$$(y_1 \ y_2) \bullet (i_1+i_2 \ i_1)^T = (y_1 \ y_2) \bullet (i_1+i_2+1 \ i_1)^T,$$

which gives us $(y_1 \ y_2) = (0 \ 1)$, i.e., the column-major layout.

When the same array is accessed in multiple nests, however, the problem of determining memory layouts program-wide becomes a complex problem (e.g., different loop nests may require different memory layouts). In Section 3, we discuss our constraint processing based solution to the problem of memory layout determination.

It is to be noted that, if a loop restructuring is applied to the nest being optimized, one can have a different data access pattern from the original one, and this can affect the memory layout selection as well. For example, if the two loops shown in Figure 2 are interchanged, the best memory layouts for arrays Q1 and Q2 would be $(0 \ 1)$ and $(1 \ -1)$, respectively.

We now briefly discuss how we handle arrays with more than two dimensions. In such cases, to define a memory layout, instead of a hyperplane vector/family, we use an ordered set of hyperplane vectors/families. For example, two data elements in a three dimensional array stored as column-major have spatial locality with respect to $(0 \ 0 \ 1)$ and $(0 \ 1 \ 0)$; that is, if they have the same indices except for the first dimension. Therefore, to represent such a layout, we use a matrix with two rows: $Y1 = (0 \ 0 \ 1)$ and $Y2 = (0 \ 1 \ 0)$. Then, the two data elements, d1 and d2, map on the same column if and only if both of the following equalities are satisfied:

$$Y_1 \bullet d_1 = Y_1 \bullet d_2 \quad \text{and} \quad Y_2 \bullet d_1 = Y_2 \bullet d_2.$$

The idea is easily generalized to higher dimensionalities as well.

## 3. Constraint Network Formulation

A constraint network (CN) can be described as a triple CN = <P,M,S>, where P is a finite set of variables, M is a list of possible values for each variable, and S is a set of constraints on P [3]. In our context, P = {Q1, Q2, …, Qz} is the set of arrays manipulated by the application code to be optimized. M, which represents the domain for variables, contains the set of memory layouts for each array (variable). Specifically, for every array Qi where $1 \leq i \leq z$, we have a set $M_i = \{h_{i1}, h_{i2}, …, h_{if(i)}\}$, which contains the hyperplane vectors that can be assumed by Qi. Here, f(i) is the number of potential layouts for array Qi. The set S, on the other hand, contains s constraints. Each $S_{ij} \in S$ contains a set of (hyperplane) pairs that capture the allowable layouts for arrays Qi and Qj from the locality viewpoint. Each pair represents the best layout choice under a given loop restructuring. In the rest of our discussion, when there is no confusion, we use the terms "array" and "variable" interchangeably. As an example, consider the following constraint network that captures layout information for a program that manipulates four different arrays (Q1,Q2,Q3,Q4):

CN = <P,M,S>, where
P={$Q_1,Q_2,Q_3,Q_4$}
M={$M_1,M_2,M_3,M_4$}, where
    $M_1$={(1 0), (0 1), (1 1)};
    $M_2$={(1 -1), (1 1)};
    $M_3$={(0 1), (1 1), (1 2)};
    $M_4$={(1 0), (0 1), (1 1)};
S={$S_{12},S_{13},S_{14},S_{23},S_{24},S_{34}$}, where
    $S_{12}$={[(1 0), (1 1)], [(0 1), (1 -1)]}
    $S_{13}$={[(1 0), (0 1)], [(0 1), (1 1)], [(1 1), (1 2)]}
    $S_{14}$={[(1 0), (1 0)], [(0 1), (0 1)]}
    $S_{23}$={[(1 1), (0 1)], [(1 -1), (1 1)]}
    $S_{24}$={[(1 0), (0 1)], [(1 1), (1 0)]}
    $S_{34}$={[(0 1), (1 0)]}.

In this constraint network, M1 indicates that array Q1 can assume three different memory layouts, represented by hyperplane vectors (1 0), (0 1), and (1 1), which correspond to row-major, column-major, and anti-diagonal layouts, respectively. Other domain sets can be interpreted in a similar fashion. S12 indicates that, as far as arrays Q1 and Q2 are concerned, there are two preferable memory layout combinations. The first combination is that Q1 has layout (1 0) and Q2 has layout (1 1), whereas the second combination is that Q1 and Q2 have layouts (0 1) and (1 -1), respectively. Note that this is similar to the situation given in Figure 2. Other constraints can be interpreted similarly. Note that, based on the way that it is encoded above, this constraint network is a binary constraint network [3] as each constraint is defined on a pair of variables. While a non-binary formulation is also possible, there are also techniques that can be used to convert non-binary formulations to binary ones. However, since it is not the main focus of this paper, we do not elaborate on this issue any further.

A solution to a constraint network problem is to select a pair from each Sij such that all the selected pairs are consistent with each other, i.e., there is no contradiction when all the members of S are considered, meaning that each array has a single memory layout. For our example above, we obtain a solution by selecting hyperplanes (1 0), (1 1), (0 1), and (1 0) for arrays Q1, Q2, Q3 and Q4, respectively. The next section discusses our strategy for finding solutions for a given constraint network (when a solution exists).

---

[2] While one can claim that we could have used (2 -2) or other similar vectors as well instead of (1 -1), this would increase the resulting data space size as some elements of the transformed data space would not be used.

**Table 1. Benchmark codes.**

| Benchmark | Brief Description | Domain Size | Data Size |
|---|---|---|---|
| Med-Im04 | medical image reconstruction | 258 | 825.55KB |
| MxM | triple matrix multiplication | 34 | 1,173.56KB |
| Radar | radar imaging | 422 | 905.28KB |
| Shape | pattern recognition and shape analysis | 656 | 1,284.06KB |
| Track | visual tracking control | 388 | 744.80KB |

## 4. Proposed Solutions

In this section, we first discuss a backtracking based solution to the problem of memory layout determination. After that, we discuss how we can shorten the solution time by enhancing the backtracking based solution with several heuristics. Before moving into the discussion of backtracking however, let us make an important definition: consistent partial instantiation.

In a constraint network, a partial instantiation of a subset of variables is an assignment to each variable from its domain. A consistent partial instantiation, on the other hand, is a partial instantiation that satisfies all the constraints that involve only the instantiated variables [3]. A backtracking algorithm traverses the state space of partial instantiations in a depth-first manner. It starts with an assignment of a variable (e.g., randomly selected) and then increases the number of partial instantiations. When it is found that no solution can exist based on the current partial instantiation, it backtracks to the previous variable instantiated, and re-instantiates it with a different value from its domain. Therefore, a backtracking algorithm has both forward (where we select the next variable and instantiate it with a value) and backward phases (where we return to the previously instantiated variable and assign a new value to it). In the rest of the paper, this backtracking based scheme is referred to as the base scheme.

The base scheme makes random decisions at several points. The first random decision is to select the next variable (array) to instantiate during the forward phase. The second random decision occurs when selecting the value (layout) with which the selected variable is instantiated (again in the forward phase). In addition, in the base scheme, when we find out that the current instantiation cannot generate a solution, we always backtrack to the previously assigned variable, which may not necessarily be the best option. One can improve these three aspects of the base scheme as follows. As for the first random decision, we replace it with an improved approach that instantiates, at each step, the variable that maximally constrains the rest of the search space. The rationale behind this is to be able to detect a dead-end as early as possible during the search. Similarly, when selecting the values to be assigned to the instantiated variable, instead of selecting a value randomly, we can select the value that maximizes the number of options available for future assignments. The rationale behind this is to increase the chances for finding a solution quickly (if one exists). Finally, we can expedite our search by backjumping, i.e., instead of backtracking to the previously instantiated value, we can backtrack further when it is beneficial to do so. This can be best explained using the following scenario. Suppose that, in the previous step, we selected the layout hid for array $Q_i$, and in the current step we selected the layout hje for array $Q_j$. If, at this point, we see that there cannot be any solution based on these assignments, the base approach returns to array $Q_i$ and assigns a new layout (say hil) to it (assuming that we tried all alternatives for $Q_j$). However, it must be noted that, if there is no constraint in the network in which both $Q_i$ and $Q_j$ (i.e., their layouts) appear together, assigning a new value (layout) to $Q_i$ would not generate a solution, as $Q_i$ is not the culprit for reaching the dead-end. Instead, backjumping skips $Q_i$ and determines an array (say $Q_k$) among the arrays that have already been instantiated that co-appears with $Q_j$ in a constraint, and assigns a new value (layout) to it (i.e., different from its current value). In this way, backjumping can prevent useless assignments and, as a result, expedite our search. Figure 3 gives an illustration that compares backtracking and backjumping. In the remainder of the paper, the solution scheme supported by these three improvements is referred to as the enhanced scheme. The next section presents experimental data for both the base and enhanced schemes. Before going into our experimental analysis though, we need to make one point clear. If a solution exists to the problem under consideration, both the base and enhanced schemes will find it. However, if multiple solutions exist, they can find different solutions.

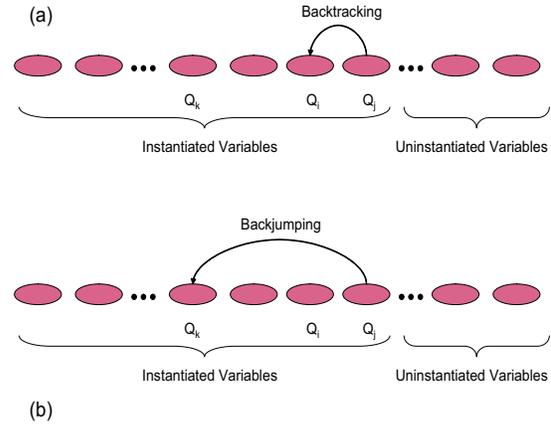

**Figure 3. (a) Backtracking. (b) Backjumping.**

**Table 2. Solution times taken by different versions.**

| Benchmark | Heuristic | Base | Enhanced |
|---|---|---|---|
| Med-Im04 | 7.14sec | 97.34sec | 12.22sec |
| MxM | 5.18sec | 36.62sec | 9.24sec |
| Radar | 11.33sec | 129.51sec | 53.81sec |
| Shape | 16.52sec | 197.17sec | 82.06sec |
| Track | 10.09sec | 155.02sec | 68.50sec |

**Table 3. Execution times achieved by different versions.**

| Benchmark | Original | Heuristic | Base | Enhanced |
|---|---|---|---|---|
| Med-Im04 | 204.27sec | 128.14sec | 82.55sec | 81.07sec |
| MxM | 69.31sec | 28.33sec | 28.33sec | 28.33sec |
| Radar | 192.44sec | 110.78sec | 83.92sec | 85.15sec |
| Shape | 233.58sec | 140.30sec | 106.45sec | 106.45sec |
| Track | 231.00sec | 127.61sec | 97.28sec | 95.30sec |

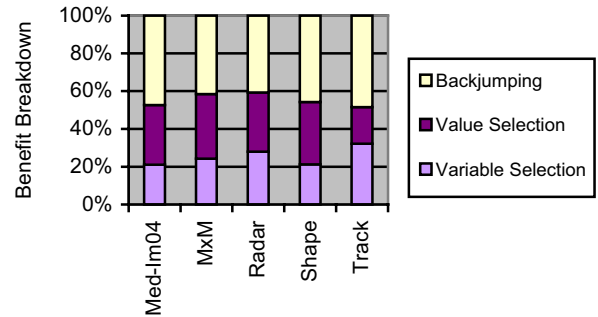

**Figure 4. Breakdown of benefits coming from the enhanced scheme.**

## 5. Empirical Analysis

In this section, we present an experimental analysis of our constraint processing based approach to the memory layout determination problem. To conduct such an analysis, we used five array-based embedded benchmarks, whose important properties are given in Table 1. The third column gives the total search space size (i.e., the sum of the domain sizes of the arrays in the corresponding application). The last column of this table gives the total data size manipulated by each application. We implemented our constraint network using C++. Excluding the libraries linked and comment lines, the network code itself is about 1700 C++ lines.

In our experimental evaluation, our focus is on two metrics: solution time and quality of solution. The first of these gives the time it takes for determining the memory layouts of arrays, and the second one gives the execution time of the resulting optimized code (or percentage improvement brought by the optimized code over the original one). To be fair in our evaluation, we also compare our approach to a previously proposed heuristic solution to memory layout optimization approach. This previous approach [9] is linear algebra based and can be summarized as follows. First, the loop nests in the program are ordered according to an importance criterion (e.g., time taken by each nest). After that, the heuristic approach processes each nest in turn, starting with the most important one (as determined by the previous step). For each loop nest being processed, it determines a good combination of loop transformation and memory layouts (for the arrays accessed by that nest). It then propagates these layouts to the second most important nest, and proceeds the same way as in the first nest except that it only determines the layouts of the arrays which are not accessed in the first nest (but accessed in the current one). In this way, it keeps propagating the memory layouts across the nests until all the layouts have been determined. Notice that, since the loop nests are ordered beforehand, this approach tends to give priority to satisfying the layout requirements of costly nests.

Our experiments have been performed using the SimpleScalar infrastructure [13]. Specifically, we modeled an embedded processor that can issue and execute two instructions in parallel. The machine configuration we use includes separate L1 instruction and data caches; each is an 8KB, 2-way set-associative with a line size of 32 bytes, and a unified 64KB L2 cache (4-way associative with a 64 bytes line size). The L1 and L2 latencies are 1 and 6 cycles respectively; and, the main memory latency is 70 cycles.

Table 2 gives the solution times for different optimized versions (in seconds) obtained on a 500MHz Sun Sparc architecture. The second, third and fourth columns give the solution times taken by the heuristic, base and enhanced schemes, respectively. We see that our base scheme takes much more time compared to the heuristic method. However, the enhanced scheme reduces these solution times dramatically, making them even comparable to those of the heuristic solution in two cases. Overall, we see that the solution times taken by our approaches are not excessive for an embedded system.

To explain how the enhanced scheme improves the solution times over the base scheme, we give in Figure 4 the percentage of reductions (in solution times) brought by each of the three enhancements discussed earlier in Section 4 (i.e., their individual contributions to the overall savings achieved by the enhanced scheme). The first enhancement is to do with the selection of the variable to instantiate next; the second one is related to the selection of the value to be assigned to the selected variable; and the last one is to employ backjumping instead of backtracking. We see the results from Figure 4 that, while most of the benefits come from backjumping, all three enhancements are very useful in general and contribute a lot to the overall reduction in solution times.

Table 3 gives the execution times for our benchmarks achieved by the original codes, heuristic approach and our constraint based approach. We see from these results that, while the heuristic solution improves over the original codes significantly (42.49% on average), the savings brought by the base and enhanced schemes are much larger: 57.17% and 57.95% on average respectively. The additional improvements are due to more comprehensive search space traversal implemented by the constraint network based approach. We also observe a small difference between the base and enhanced schemes. This difference is due to the fact that these two schemes can find "different solutions" if there are multiple solutions to the underlying network (as in the case of Med-Im04, Radar, and Track).

## 6. Conclusions and Future Directions

Recent years have witnessed, from both embedded system community and scientific computing community, a large number of studies targeting at improving data cache behavior of array based codes. Data transformations in particular have been found attractive as they do not have the drawbacks of commonly used loop transformations. This paper presents a novel constraint processing based approach to data transformations, where the problem of memory layout determination is captured as finding solutions in a constraint network. Our empirical analysis shows that the proposed approach is very effective in practice, and further enhancements are possible to expedite the search in the constraint graph.

We plan to extend this work in two directions. First, we would like to give weights to constraints. This will help us distinguish between different solutions to a given network. Second, we would like to expand our constraint network formulation to accommodate dynamic memory layouts, i.e., the layouts that can change during execution based on the requirements of the different segments of the program.